\documentclass{article}
\usepackage{smc2025}
\usepackage[caption=false, font=footnotesize]{subfig}
\usepackage{paralist}
\usepackage[figure,table]{hypcap}

\usepackage[whole]{bxcjkjatype}

\usepackage{alphabeta}
\usepackage{arabtex}
\usepackage[LFE,LAE,LGR,T2A,T1]{fontenc}
\usepackage[greek, russian, main=english]{babel}

\def\papertitle{Computational Extraction of Intonation and Tuning Systems from Multiple Microtonal Monophonic Vocal Recordings with Diverse Modes}

\author[1]{\mbox{\firstname{Sepideh}\lastname{Shafiei}}}
\author[2]{\mbox{\firstname{Shapour}\lastname{Hakam}}}

\affil[1,2]{\institution{Cu Test Inc.}\city{Berkeley, California}\country{USA}\affiliationtype{Company}}
\completesetup

\title{\papertitle}
\begin{document}
\capstartfalse
\maketitle

\capstarttrue

\begin{center}
\small\itshape
Author’s version (preprint). \par
This paper appears in the \emph{Proceedings of the Sound and Music Computing Conference (SMC 2025)}. \par
© 2025 The Authors. Licensed under \href{https://creativecommons.org/licenses/by/4.0/}{CC BY 4.0}. \par
Please cite the published version: \url{https://zenodo.org/records/15838081}
\end{center}

	\begin{abstract}
		This paper presents a computational methodology for analyzing intonation and deriving tuning systems in microtonal oral traditions, utilizing pitch histograms, Dynamic Time Warping (DTW), and optimization techniques, with a case study on a complete repertoire performed by a master of Iranian Classical Vocal Music (145 pieces). Pitch frequencies are extracted directly from vocal performances, and while alignment with MIDI notes is not a standard practice in our approach, we incorporate it where available, using DTW to refine interval analysis. By modeling intonation variations across multiple recordings, we derive structured tuning frameworks that capture both the flexibility of performance and the underlying systematic tendencies. Optimization techniques are applied to align intervals across the oral tradition repertoire, capturing the specific tunings and modal structures involved. Our methodology highlights the potential of computational techniques in advancing musicological and ethnomusicological research, offering a data-driven approach to defining tuning systems in microtonal vocal traditions.
		
	\end{abstract}

\section{Introduction}

Understanding the exact musical intervals performed in various traditions is crucial for several reasons. First, it offers deeper insights into the cultural and theoretical foundations of a musical system, highlighting the unique characteristics that distinguish different traditions. Precise interval analysis can uncover the microtonal nuances often overlooked in conventional studies—an essential aspect of oral traditions, where music is transmitted through performance rather than notation. Additionally, analyzing performed musical intervals provides insights into intonation as an aspect of the performer’s style, capturing the expressive choices that shape musical traditions.

Accurate interval identification also plays a vital role in preserving and documenting musical heritage, ensuring that the subtleties of performance practices are not lost over time. This preservation is crucial for maintaining the authenticity and continuity of cultural expressions across generations. Furthermore, precise interval analysis has practical applications in music education, performance practice, and digital archiving. In music education, a detailed understanding of intervals enhances the teaching and learning process, helping students grasp the nuances of different musical systems and their theoretical underpinnings. In performance practice, it enables musicians to reproduce traditional music with greater fidelity, respecting the intricate details of intervallic structures. For digital archiving, precise computational documentation ensures that recordings and notations of traditional music remain accurate, comprehensive, and accessible for future research and exploration.

By employing computational methods such as pitch histograms, Dynamic Time Warping (DTW), and optimization techniques, researchers can achieve a more precise and systematic analysis of intonation and tuning. These methods allow for the direct extraction of pitch frequencies from multiple vocal performances, enabling the detection of microtonal intervals without reliance on predefined tuning systems. While MIDI note alignment is not central to our approach, we incorporate it where available using DTW to refine interval analysis. To demonstrate this methodology, we apply it to a comprehensive repertoire of 145 pieces performed by a master of Iranian Classical Vocal Music. This dataset is particularly well-suited for analysis due to its extensive coverage of traditional modes, rich microtonal details, and high-quality performances that exemplify the stylistic and intonational practices of the tradition. Optimization techniques further refine this process by systematically modeling intervallic relationships across the repertoire, capturing both intonational fluidity and underlying tuning structures.


\section{Choice of Repertoire}

To demonstrate our methodology and examine its results, we have selected a comprehensive repertoire of 145 pieces from Iranian classical music, performed by one of the twentieth century's masters, Karimi. These performances have been publicly recorded and transcribed by the prominent ethnomusicologist Masʿudiye \cite{M1}. The transcriptions were later digitized in Finale, with pitch bends incorporated into the MIDI playback. This data has been compiled into a database of Iranian classical music as part of the first author's dissertation \cite{Sh1}. 

Iranian vocalists are among the few musicians who continue to preserve and transmit music through oral tradition, relying on master-to-student transmission rather than musical notation. In the foreword to Maʿrufi’s radif, Barkeshli emphasizes that the human voice provides a more authentic reference for Iranian musical intervals than musical instruments. Unlike instruments, which may introduce structural tuning constraints, the human voice remains unbounded, allowing for greater microtonal flexibility \cite{Ma1}. According to Barkeshli, measuring the Persian scale through vocal performances offers a more genuine representation of traditional intonation. Furthermore, in Iranian classical music, the human voice serves as the primary pitch reference, with instruments—especially unfretted ones like the kamancheh and ʿud—seeking to closely follow vocal intonation patterns.

\begin{figure}
\centerline{\framebox{
\includegraphics[width=0.9\columnwidth]{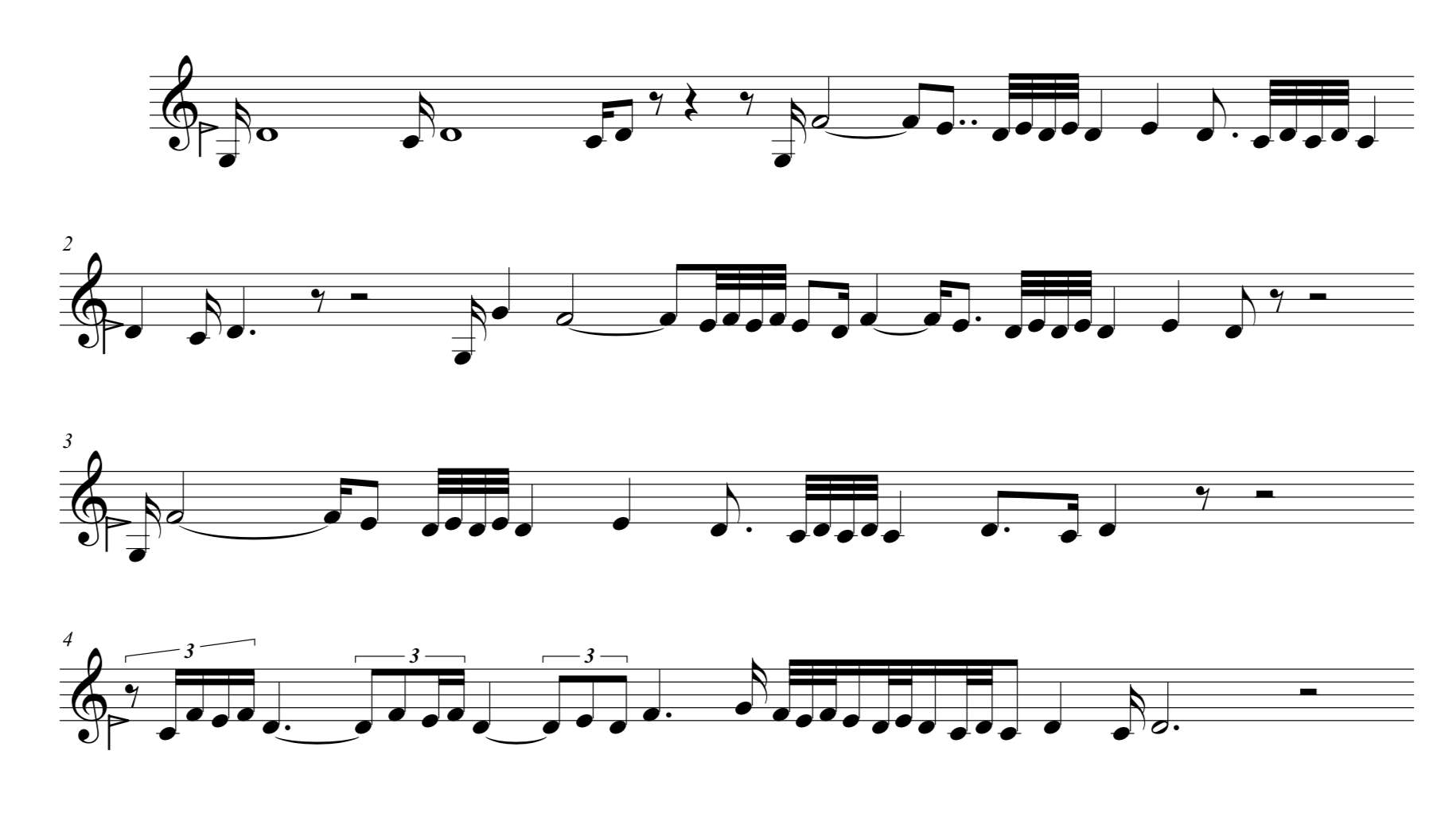}}}
\caption{Transcription of Example 1}
\label{fig:daramad}
\end{figure}

It is important to note that while this study employs Karimi’s repertoire as a case study, the proposed computational methodology is repertoire-independent and can be easily adapted to analyze other musical traditions with minimal modification. The chosen repertoire serves as a representative example due to its comprehensive coverage of traditional modes, rich microtonal details, and the high-level artistry of its performance. Moreover, the fluidity of intonation in these performances provides an ideal dataset for studying how microtonal intervals are realized dynamically in practice, rather than being confined to fixed theoretical models. To illustrate the methodology, we have selected the first piece of the chosen repertoire, referred to as Example 1, throughout the paper. Figure \ref{fig:daramad} presents its transcription. In this notation, the p-shaped sign on the note E represents \emph{koron}, a microtonal modification in Iranian Classical Music that lowers the pitch by approximately a quarter-tone, though its exact tuning varies depending on performance context.

In standard MIDI convention, each pitch is represented by an integer; for example, 55 corresponds to G2, 56 to G\#2, and 57 to A2. To accommodate microtonal intervals such as \emph{sori} (half-sharp) and \emph{koron} (half-flat), we extend this system by doubling the MIDI resolution. Specifically, we multiply each MIDI pitch number by two, enabling us to insert intermediary microtonal steps. For example, in our system, 110 represents G2, 111 corresponds to G2-half-sharp (\emph{sori}), 112 to G\#2, 113 to A2-half-flat (\emph{koron}), and 114 to A2.

\section{Pitch Recognition}

There are various algorithms for pitch recognition, all of which rely on fundamental frequency estimation to determine pitch. The fundamental frequency (F0), the lowest frequency component of a sound wave, corresponds to its most dominant perceived pitch. Monophonic voice pitch estimation algorithms have been extensively studied in Music Information Retrieval (MIR) literature. Gómez et al. evaluated two leading algorithms, CREPE and pYIN, for monophonic vocal pitch recognition \cite{GBBC-9}. Their evaluation on the iKala dataset reported 91\% Raw Pitch Accuracy for pYIN and 90.5\% for CREPE. Another notable method, SPICE \cite{BGSpice9-a}, employs a self-supervised learning technique for pitch estimation.

For this study, we selected pYIN due to its strong performance and extensive validation in MIR research. To minimize errors, we performed manual octave-error corrections when necessary. The output from pYIN is stored in \emph{.csv} format, allowing for easy substitution with alternative pitch recognition algorithms for experimentation or future improvements.

To generate the pitch file, we used Sonic Annotator with the following parameters: a step size of 256, a block size of 2048, low amplitude suppression of 0.1, onset sensitivity of 0.7, prune threshold of 0.1, and a threshold distribution of 2. All audio files were sampled at 44.1 kHz. Pitch values extracted in Hertz (Hz) were converted to the cents system, where each octave equals 1200 cents. The interval between two notes with frequencies $f_1$ and $f_2$ in cents is given by the following formula:

\[
\text{cents}(f_1, f_2) = 1200 \cdot \log_2\left(\frac{f_2}{f_1}\right)
\]

While the fixed block size provides a consistent basis for our analyses, we acknowledge that certain aspects of microtonal vocal performance—particularly rapid expressive gestures—might benefit from multiresolution or adaptive time-frequency representations. Exploring such approaches remains a promising direction for future development.



\section{Interval Estimation}

\subsection{Pitch Histogram}

\par Pitch histograms are a valuable tool in Music Information Retrieval (MIR) for visualizing and analyzing musical intervals. They provide a statistical representation of pitch frequencies within a piece, revealing patterns and structures that characterize specific musical traditions. This technique is particularly valuable for handling large datasets and is applicable across various musical genres, including microtonal music in oral traditions. 

Histograms have been applied extensively to non-Western and microtonal musical traditions, providing insights into unique interval patterns and tuning systems. Serra used histograms to analyze the maqam system in Arab music, identifying specific microtonal intervals that define this tradition  \cite{Serra11}. Similarly, Srinivasamurthy et al.  utilized histograms to study raga structures in Indian classical music, revealing distinct interval patterns that differentiate ragas from each other \cite{Srini}. Barış Bozkurt has significantly contributed to the use of pitch histograms in the analysis of Turkish makam music \cite{Boz8}. Bozkurt's work emphasizes the importance of computational methods for analyzing the microtonal intervals characteristic of makam music, providing a deeper understanding of its intricate tuning systems. 

The accuracy and resolution of pitch extraction algorithms are crucial for reliable histogram analysis. Salamon and Gómez emphasized the importance of high-quality pitch extraction for constructing accurate histograms in their work on melody extraction from polyphonic music  \cite{SG}. Statistical methods for analyzing histograms, such as clustering and classification algorithms, further enhance the robustness of interval analysis. Peeters integrated spectral analysis with histogram data to study the timbral characteristics of pitches, offering a richer analysis of musical content \cite{Peet}. Machine learning algorithms, as used by Berg-Kirkpatrick et al. can classify and cluster interval patterns, uncovering deeper structural insights into musical traditions \cite{Berg}. Pitch histograms has also been used in analysis of pitch drift in unaccompanied solo singing \cite{Sh4}. 

\begin{figure}
 \centerline{\framebox{
 \includegraphics[width=0.9\columnwidth]{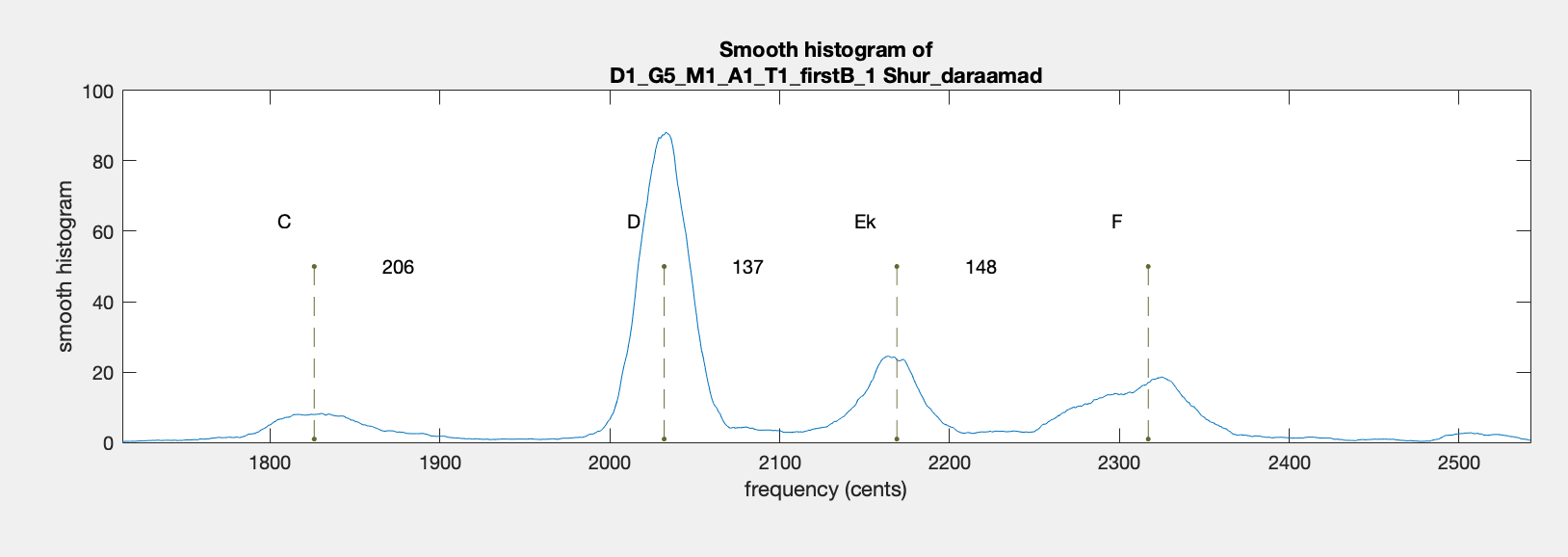}}}
 \caption{Pitch Histogram of Example 1}
 \label{fig:audio histo}
\end{figure}

\begin{figure}
 \centerline{\framebox{
 \includegraphics[width=0.9\columnwidth]{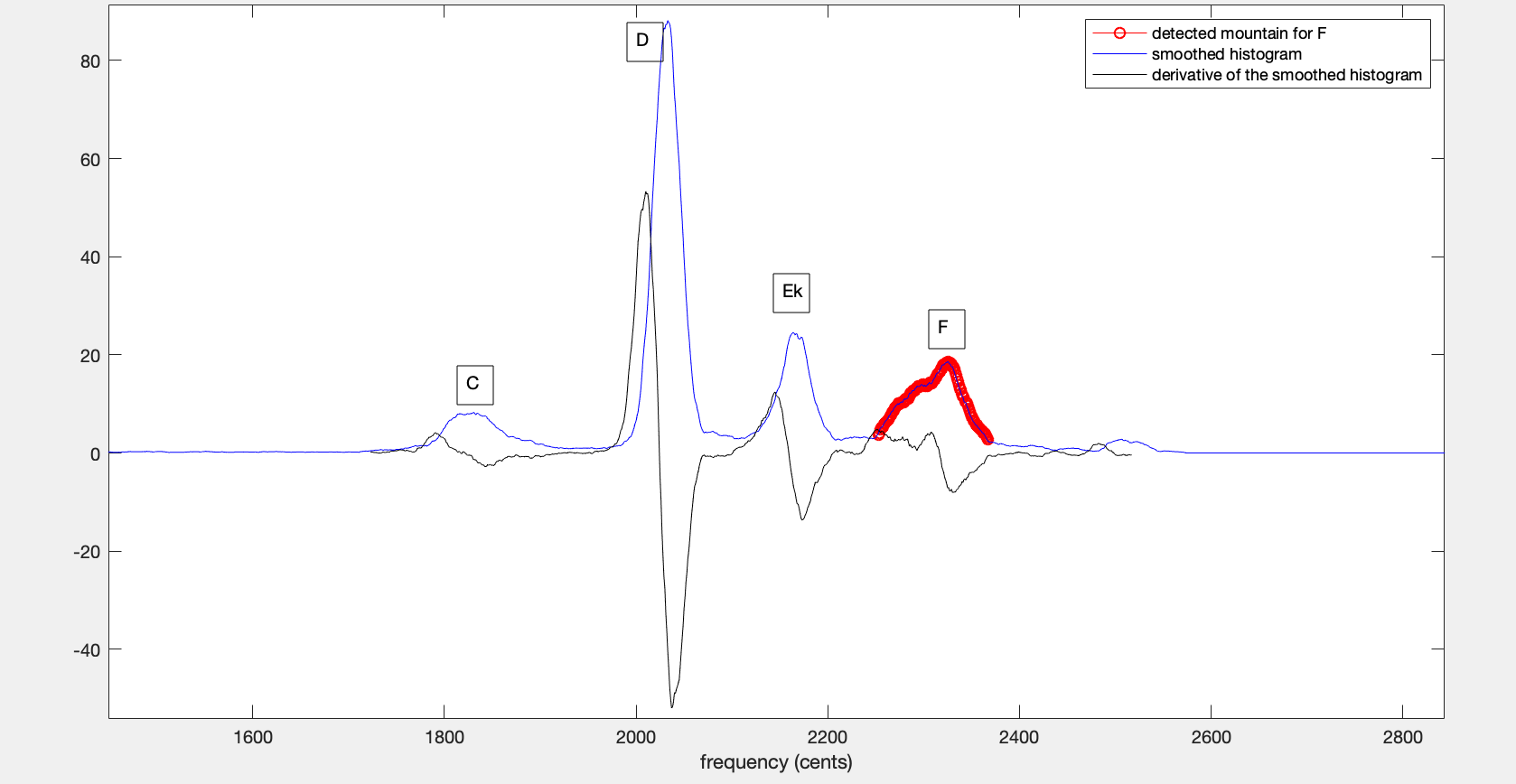}}}
 \caption{Determining the Range of note F in Example 1}
 \label{fig:F range}
\end{figure}

Figure \ref{fig:audio histo} presents the pitch histogram of the audio recording of Example 1. To identify the peaks, we first determine the range of each mountain. This process is challenging, as peak boundaries are not always clearly defined. We employ different methods based on both the shape of the mountain and the derivative of the smoothed histogram. Tracking sign changes in the derivative curve provides an indication of peak boundaries. Figure \ref{fig:F range} illustrates this process for determining the range of the mountain corresponding to the note F in Example 1. In Figures~\ref{fig:audio histo} and~\ref{fig:F range}, we only annotate peaks that are supported by a sufficient number of data points to ensure reliable estimation. Although a small peak corresponding to G is present, its low height suggests limited data coverage, making accurate identification and interpretation less robust. For this reason, such minor peaks are not labeled or further analyzed in this study.

 \par

As shown in Figure \ref{fig:audio histo}, some notes exhibit greater pitch flexibility. Understanding the cause of these wider peaks requires further investigation, which will be discussed in Section \ref{Types of Peaks}. Once the range of each peak is determined, we approximate it using a tilted Gaussian curve to refine peak detection. Selecting appropriate peaks is critical, as they serve as key reference points for further analysis. These peaks are then used to identify the primary reference pitch in the audio signal, facilitating the automatic extraction of the scale from the performance.
The tilted Gaussian curve model is defined by the following equation:

\begin{equation}
{y=c}_1+c_2x+c_3e^{-(x-c_4)^2/c_5}
\end{equation}

We apply a non-linear curve fitting method to estimate the parameters $c_1,\ldots,\ c_5$ ensuring an optimal fit to the data.  Section \ref{Types of Peaks} further explores the different types of histogram peaks observed in the selected repertoire.

\subsection{Typology of the Observed Peaks}
\label{Types of Peaks}

We use the peaks of the audio histograms to determine the performed musical intervals. In this process, it is crucial to consider the shape of each mountain and its peak, as these variations provide insights into intonation and pitch behavior within the repertoire. Our analysis reveals several distinct types of peak structures, each with different implications for pitch estimation.

\par \textbf{I. Well-Defined Gaussian Peaks} 

\begin{figure}
 \centerline{\framebox{
 \includegraphics[width=0.9\columnwidth]{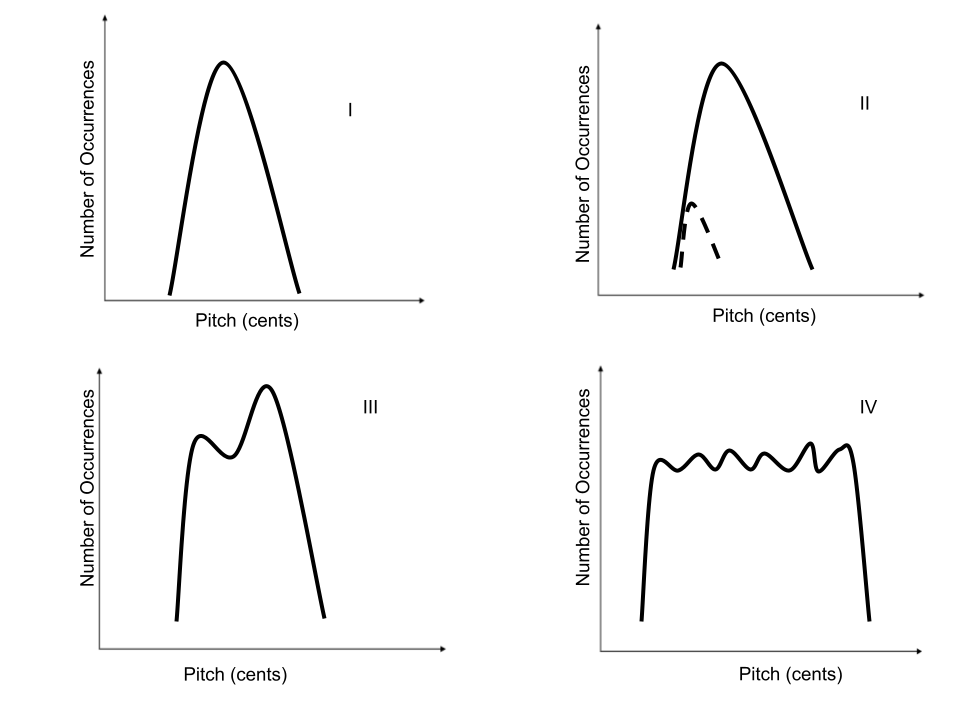}}}
 \caption{Types of Pitch Histogram Peaks}
 \label{fig:Peaks}
\end{figure}

The ideal type of peak follows a Gaussian distribution with minimal error. In this case, the peak is clear and well-defined, making it relatively easy to identify. When comparing the audio histogram to the MIDI reference, the peak in the audio histogram corresponds directly to a single MIDI note. Figure \ref{fig:Peaks}-I illustrates the general shape of this type.

\textbf{II. Hidden Secondary Peaks}

This type appears similar to the first, but when comparing audio and MIDI histograms, we observe that two MIDI notes fall within the range of this mountain. Typically, one peak is prominent, while the second is much smaller, occurring within 50 cents of the main peak. The secondary peak is often embedded within the larger mountain, making it difficult to detect unless examined in detail.
If we analyze only the audio histogram, this structure resembles Type I. However, when we consider transcription data, a small hidden peak becomes evident. In some cases, this secondary peak creates a visible bump. This type of peak often appears in variable notes within a piece. Figure \ref{fig:Peaks}-II illustrates this structure.

\textbf{III. Double-Peaked Mountains}
Some mountains exhibit two distinct peaks of either equal or different heights. This presents a challenge:
\begin{itemize}
\item If the two peaks correspond to separate notes in MIDI, they must be treated as two distinct peaks.
\item If the mountain represents a single note, we must determine whether to take the higher peak or calculate a peak midway between the two using Gaussian peak fitting.
\end{itemize}

\textbf{IV. Flattened or Curved Peaks}
Some peaks appear flat or curvy on top, extending over a wide interval of 30 cents or more. In these cases, we must determine whether the peak corresponds to a single note or two separate notes and investigate the reason for its unusually broad range.
Figure \ref{fig:Peaks}-IV illustrates this peak shape.

Other mountain shapes are often combinations of these four primary types. In Types II, III, and IV, there is no definitive rule for selecting the best peak. The interpretation must be considered case by case, based on the musical context and the underlying performance characteristics that shaped the histogram.

\section{Alignment of Pitch and MIDI}\label{alignment}

Parallel to the audio we have made a table corresponding to the MIDI file which contains the MIDI note and the duration of each note based on the transcription. The problem of MIDI to audio matching has a long history in the Music Information Retrieval community. There are different algorithms for comparing and matching MIDI and audio sequences. Ewert et al. mention three different approaches for the alignment problem: Dynamic Time Warping  (DTW), a recursive version of Smith-Waterman algorithm , and partial matching \cite{Ewert}. 

DTW is one of the most common algorithms that is used to compare two given time series in MIR and speech recognition. Müller et al. demonstrated the use of DTW in synchronizing music performances with corresponding scores, enabling accurate comparisons and analyses \cite{muler}. Raffel and Ellis applied DTW to align symbolic music representations with audio recordings, facilitating tasks such as automatic transcription and performance analysis. DTW's ability to handle variations in timing makes it an invaluable tool for studying the nuances of musical performances in oral traditions \cite{rafael}.

For audio-to-MIDI alignment, several DTW-based approaches exist. In our study, since we are working exclusively with monophonic voice, and the pYIN pitch recognition algorithm performs with high accuracy on our dataset, we apply DTW directly to align the extracted $F_0$ curve with the MIDI transcription.

\begin{figure}
 \centerline{\framebox{
 \includegraphics[width=0.9\columnwidth]{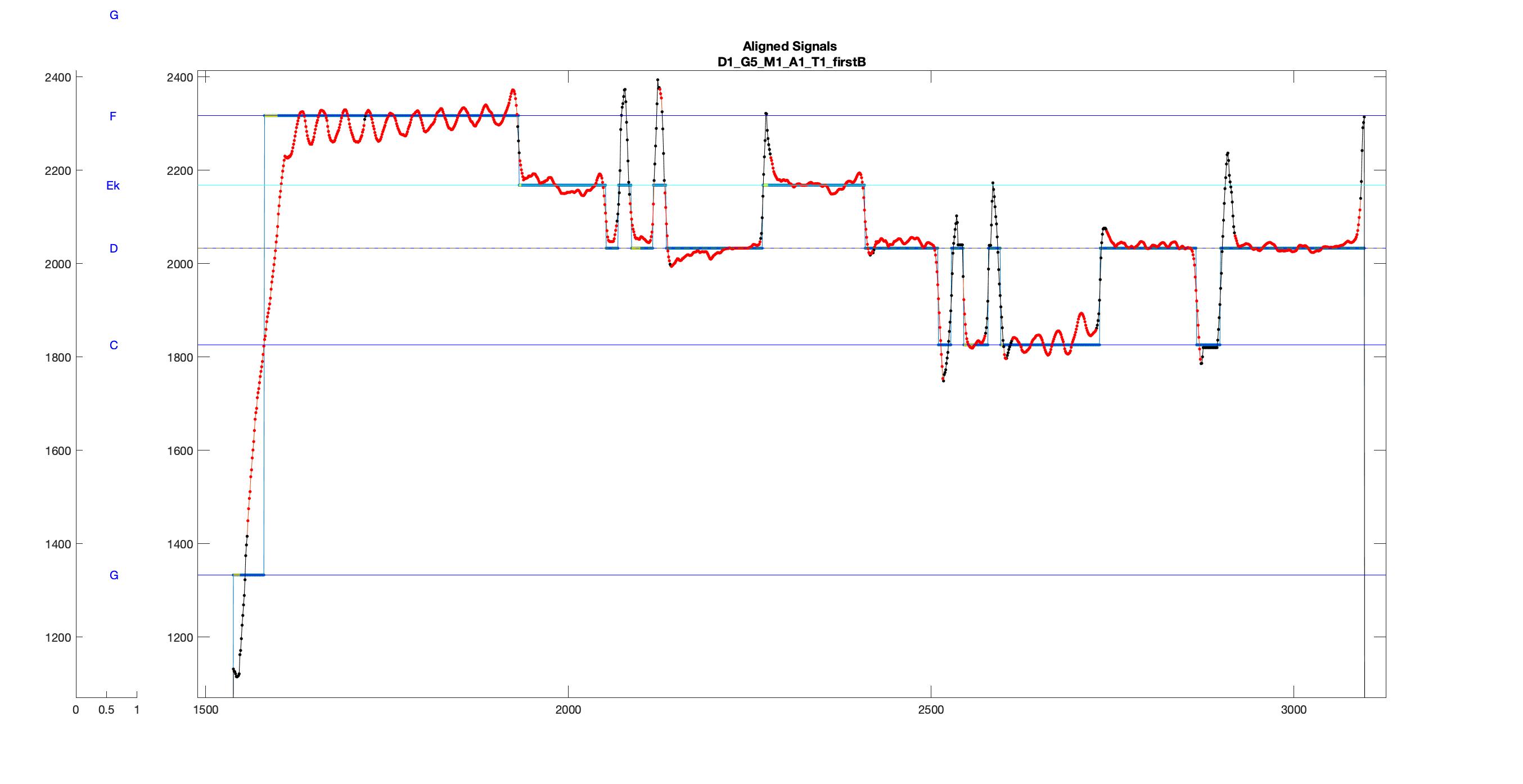}}}
 \caption{Aligned Pitch and MIDI}
 \label{fig:aligned}
\end{figure}

Figure \ref{fig:aligned} illustrates the aligned MIDI and pitch diagrams for a phrase in Example 1.

\section{Histogram of Notes}\label{noteHisto}

This method is designed to verify the results obtained from histogram peak detection, particularly in cases where the audio histogram of a piece contains unclear peaks. It is especially useful for resolving Types II, III, and IV peaks, as described in Section \ref{Types of Peaks}.

This approach leverages audio-to-transcription alignment to improve peak detection accuracy. After aligning MIDI and pitch data (as detailed in Section \ref{alignment}), we construct individual histograms for each note. Specifically, we identify all frequency points in the audio that correspond to a given MIDI note and then compute a histogram of those frequencies.

\begin{figure}
 \centerline{\framebox{
 \includegraphics[width=0.9\columnwidth]{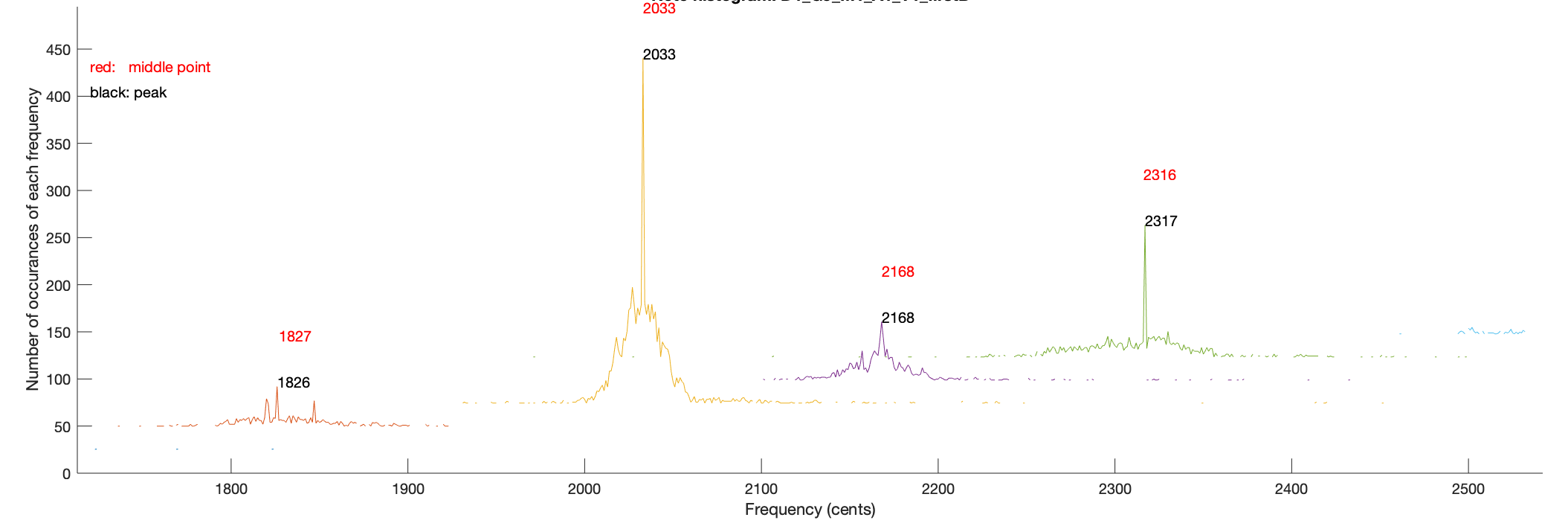}}}
 \caption{Note Histogram of Example 1}
 \label{fig:Notehisto}
\end{figure}

Figure \ref{fig:Notehisto} presents the histogram of each note for Example 1, where each peak represents the distribution of frequencies associated with a particular MIDI note.

\section{Computational Analysis of Intonation across the Repertoire}\label{Shur}

\begin{figure}
 \centerline{\framebox{
 \includegraphics[width=0.9\columnwidth]{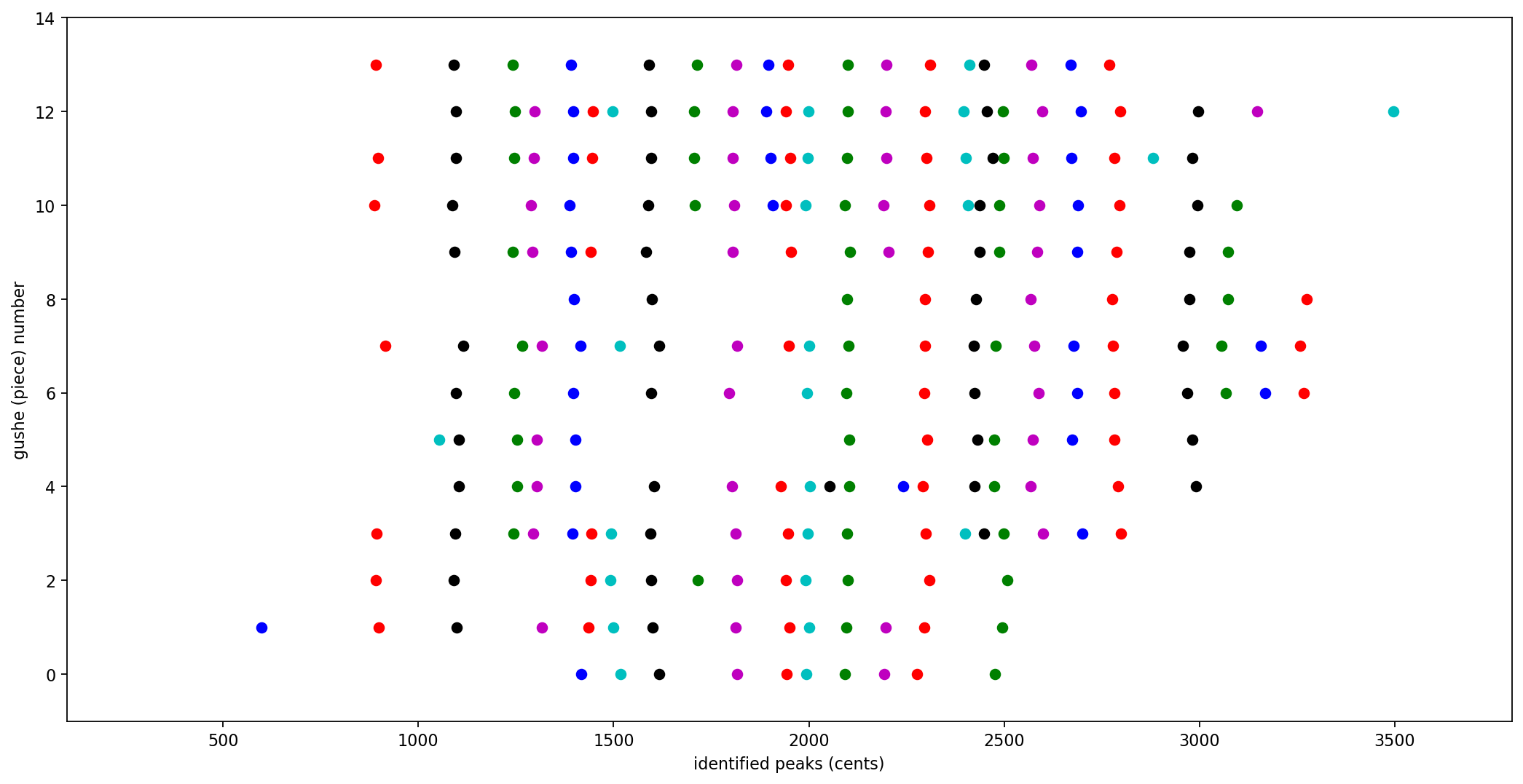}}}
 \caption{All the Observed Peaks for the first 14 Pieces of the Repertoire}
 \label{shurdots}
\end{figure}

After identifying all the peaks in the pitch histograms, we compile them into a table, where each row represents the peaks corresponding to a specific piece. For the selected repertoire, we visualize the first category of pieces (shur) in Figure  \ref{shurdots}, while the full dataset consists of 145 gushes. The complete tables are available in the corresponding GitHub repository for reference.

In the selected repertoire, the reference pitch is fluid, meaning that there is no fixed fundamental pitch across all pieces, and the intervals between notes can vary depending on the performance. For example, the interval between F and G might be 198 cents in one piece but 210 cents in another. This variability makes it challenging to align multiple pieces. 

To further illustrate the diversity of tuning behaviors in the dataset, we examined several examples across different \emph{dastgāh}s. In particular, we compared the intervals derived from audio performances with those found in musicological transcriptions of the same pieces. We observed notable inconsistencies, especially in the treatment of neutral intervals. For example, what is labeled as a "neutral second" in some cases may span anywhere from approximately 120 to 180 cents. In certain contexts, an interval of 180 cents has been labeled a neutral second, whereas in others it has been considered a major second. Similarly, intervals around 120 cents have been interpreted either as narrow neutral seconds or as minor seconds, depending on the transcriber’s theoretical framework. These discrepancies highlight both the flexibility of intonation in performance and the lack of consensus in interval classification, reinforcing the typology described in Section~\ref{Types of Peaks} and the need for data-driven representations.

To address this, we first approximate the alignment of all pieces, forming columns of corresponding notes. However, their placement will not initially be optimal. We iteratively refine their alignment by shifting the notes of each piece up or down while minimizing the total sum of standard deviations across corresponding notes, that is explained in Section \ref{optim}. 

This approach ensures that the precise intervallic relationships within each piece are maintained while allowing for a best-matching tuning system to emerge from the dataset. 

\subsection{Optimization Process for Aligning Pitch Rows in the Tuning System} \label{optim}

To achieve an optimal alignment of pitch intervals across the repertoire, we employ an iterative optimization algorithm designed to minimize variations in pitch placement. This process ensures that the standard deviation across corresponding notes in different pieces is minimized, resulting in a more coherent and representative tuning system.

The optimization follows a greedy iterative approach, where each row, representing a specific piece in the pitch matrix, is adjusted iteratively, and its impact on the overall cost function is evaluated. The algorithm runs for $n$ iterations; however, an early stopping criterion is implemented. If no improvement is observed over $l$ consecutive steps, the process terminates to prevent unnecessary computations.

The core of this optimization consists of two sequential adjustment steps: forward and backward modifications. In the forward adjustment phase, each row in the pitch matrix is modified iteratively by increasing nonzero elements by one, effectively shifting pitch values upward. After each adjustment, the cost function is computed both before and after the modification. If the adjustment reduces the cost function, it is retained; otherwise, the modification is reverted.

Following the forward adjustment, a backward adjustment phase is performed in which each row is decremented by one, shifting pitch values downward. As in the forward phase, the cost function is evaluated before and after the adjustment, and any modification that increases the cost is reverted to preserve the optimal placement.

The cost function used in this process quantifies the total deviation across the tuning system. At each step, the cost is recorded to track progress and monitor convergence. If there is no improvement in the cost function over $l$ consecutive iterations, the optimization process is terminated early.

\begin{figure}
 \centerline{\framebox{
 \includegraphics[width=0.9\columnwidth]{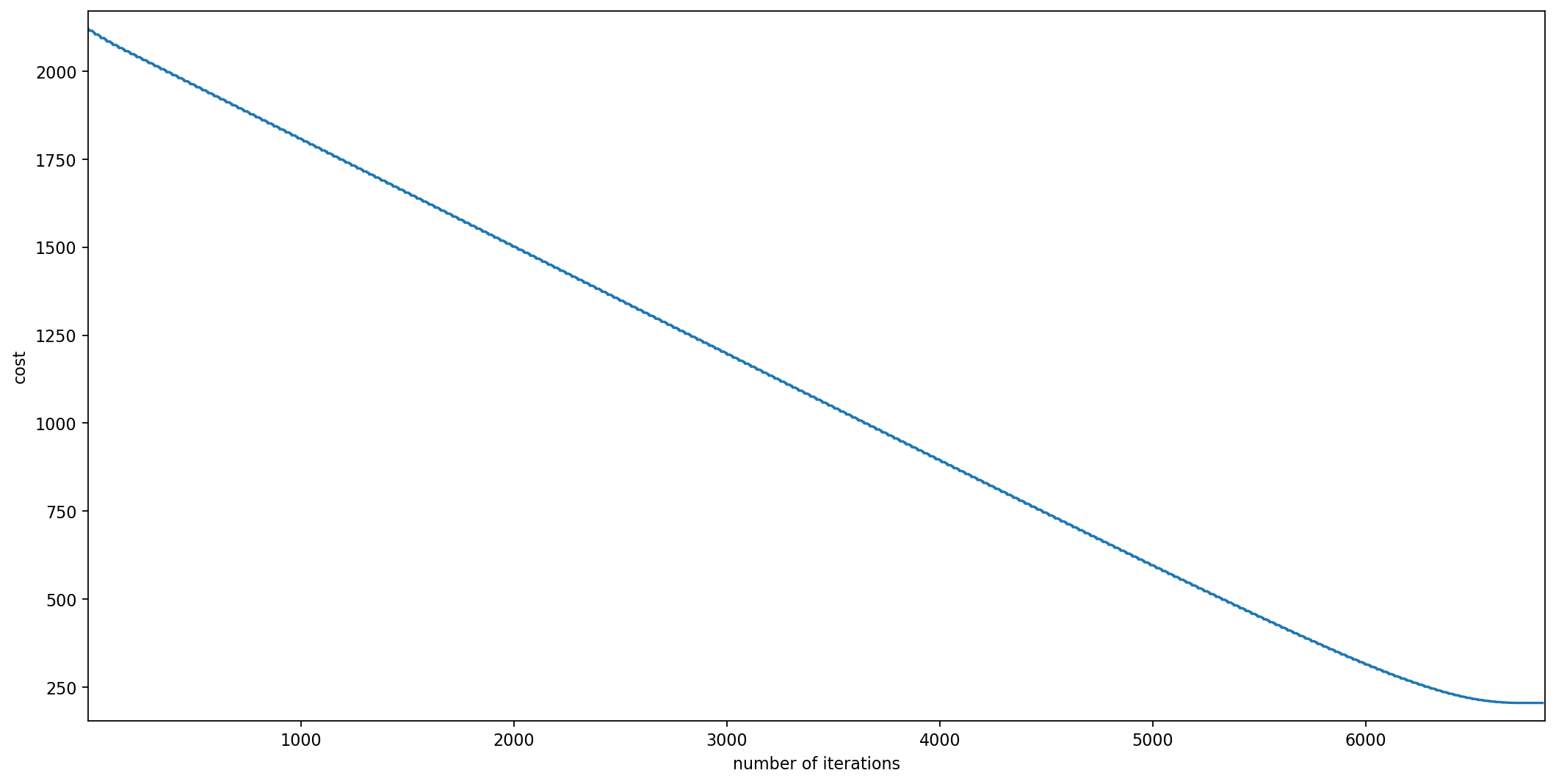}}}
 \caption{Evolution of the Cost Function During the Optimization Process}
 \label{cost}
\end{figure}

This iterative approach allows the algorithm to refine the placement of each piece within the tuning system, ensuring that the overall tuning variability is minimized while preserving the structural integrity of individual performances. The combination of forward and backward adjustments ensures that the optimization explores both increasing and decreasing pitch adjustments, leading to a well-balanced alignment. The early stopping mechanism further improves efficiency by preventing excessive computations when no further improvements are possible. Figure \ref{cost} illustrates the evolution of the cost function throughout the optimization process.

The current optimization procedure is based on a greedy iterative search that ensures monotonic descent of the cost function. This approach offers interpretability and low computational overhead, but we acknowledge that its local, discrete adjustments may limit the exploration of the solution space and lead to premature convergence. As such, this method is viewed as a baseline. In future work, we plan to explore gradient-based and probabilistic optimization strategies, which may offer more robust convergence properties and improve tuning alignment, particularly in cases involving sparse or unevenly distributed data.

\subsection{Derived Tuning and Visualization of Intonation Patterns}

Once the optimal positioning for each piece is determined within the cumulative pitch rows (Figure \ref{shurdots}), we compute the average frequency of each note across the repertoire. This provides a tuning reference derived from the complete performance dataset.

\begin{figure}
 \centerline{\framebox{
 \includegraphics[width=0.9\columnwidth]{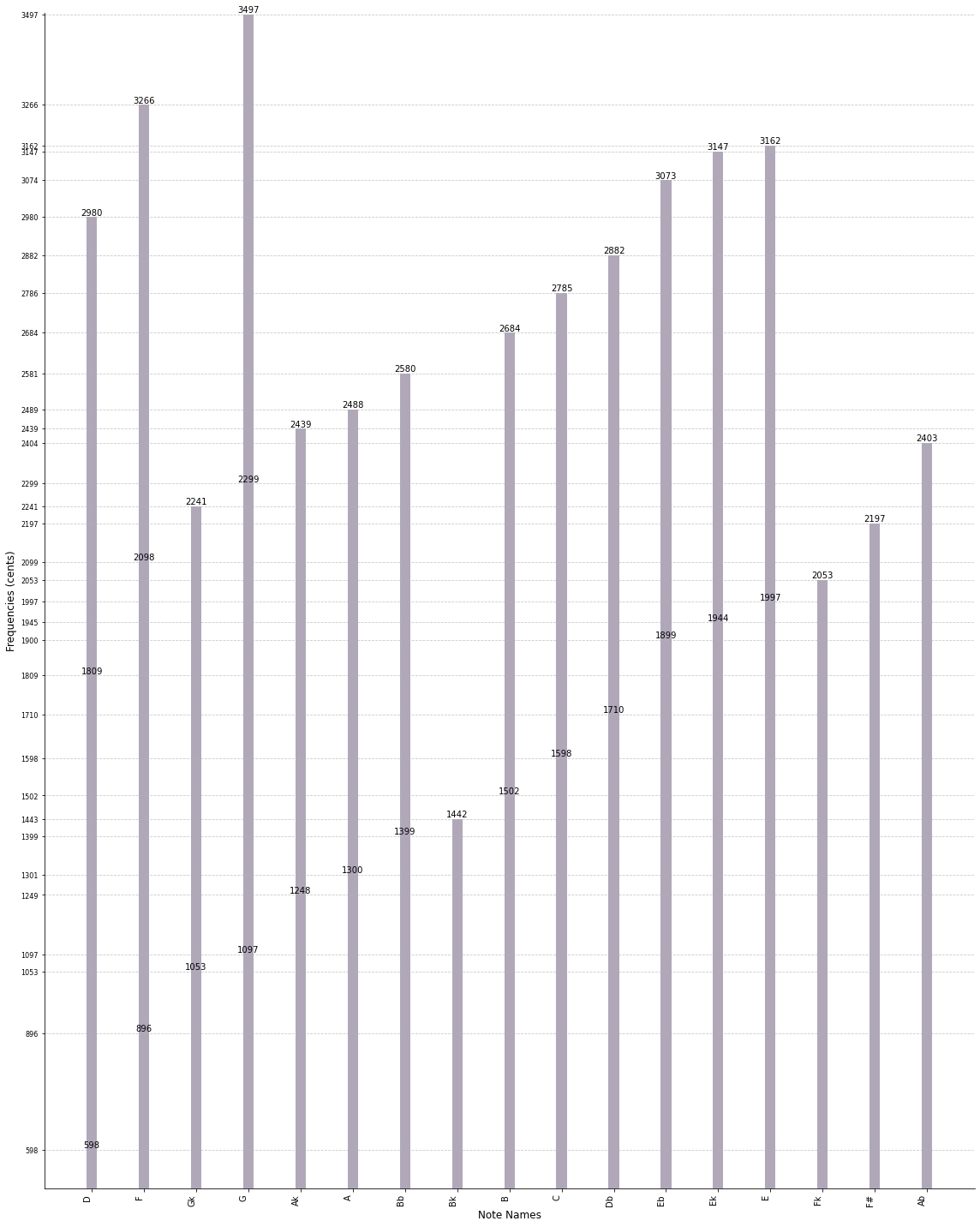}}}
 \caption{Tuning Extracted from Dastgāh of shur}
 \label{fig:imageshur}
\end{figure}

Figure \ref{fig:imageshur} shows the final tuning extracted from the multiple pieces in shur. It is not intended as a prescriptive or canonical tuning system for this dastgāh, but rather as a descriptive statistical summary of how pitches are realized across multiple pieces. While some intervals may align with theoretical expectations or commonly used tunings, others reflect the fluidity and expressive variation inherent in this repertoire. Therefore, this tuning should be understood as a performance-informed approximation, not a fixed template.


\section{Future Directions}

The methodology introduced in this study opens several promising avenues for future work. One natural extension is to apply the analysis across a broader range of vocalists within the same genre, in order to uncover stylistic characteristics of intonation that may distinguish individual performers, lineages, or pedagogical traditions. By systematically comparing pitch distributions and tuning profiles across singers, it may be possible to identify performer-specific tendencies or regionally influenced variants of modal interpretation.

In addition to stylistic analysis, this approach can also be used to explore the nuanced realizations of ornamentation and intervallic ambiguity that are central to microtonal vocal traditions. For instance, by examining how certain intervals (such as neutral seconds) are treated across different contexts, we may gain insight into their expressive functions and contextual elasticity.

\begin{acknowledgments}

This work originated as part of the first author's dissertation under the supervision of Prof. Blum at the Graduate Center, City University of New York \cite{Sh1}. The code and dataset used in this study are publicly available at \url{https://github.com/SepiSha/radifToolBox}.
\end{acknowledgments}

\bibliography{ISMIRtemplate}


\end{document}